\documentclass[12pt,showpacs,aps,pre,preprint]{revtex4}

\usepackage{amsmath}
\usepackage{mathrsfs}
\usepackage{bm}
\usepackage[pdftex]{hyperref}
\usepackage{graphicx}
\usepackage{subfigure}
\begin{document}
\begin{titlepage}
	
\title{Mathieu functions approach to bidimensional scattering by dielectric elliptical cylinders}
\author{E. Cojocaru}
\affiliation{Department of Theoretical Physics, IFIN HH, Bucharest-Magurele MG-6, Romania}

\email{ecojocaru@theory.nipne.ro}

\date{today}

\begin{abstract}
Two-dimensional scattering by homogeneous and layered dielectric elliptical cylinders is analyzed following an analytical approach using Mathieu functions. Closed-form relations for the expansion coefficients of the resulting electric field in the vicinity of the scatterer are provided. Numerical examples show the focalizing effect of dielectric elliptical cylinders illuminated normally
to the axis. The influence of the confocal dielectric cover on the resulting scattered field is envisaged.
\end{abstract}


\maketitle
\end{titlepage}
\section{Introduction}
The problem of wave scattering by an elliptical cylinder has been known for some time \cite{re1,re2}. Exact solutions can be obtained by separating the wave equation in elliptic cylindrical coordinates and constructing exact eigenfunction expansions. But these exact solutions have been of limited practical value due to problems associated with the computation of the corresponding eigenfunctions, i.e. angular and radial Mathieu functions.

Recently, a new efficient method for computing these functions was presented \cite{re3}, the method being implemented in \textsc{matlab} \cite{re4}. To extend the applicability of the new computational method, we derive in this paper the eigenfunctions for two-dimensional (2D) scattering by dielectric elliptical cylinders. Closed-form relations for the normally incident plane wave scattering by homogeneous and layered elliptical cylinders are provided.

\section{Wave equation in elliptic cylindrical coordinates}
\begin{figure}[h]
\begin{center} {\includegraphics [height=7.5cm]{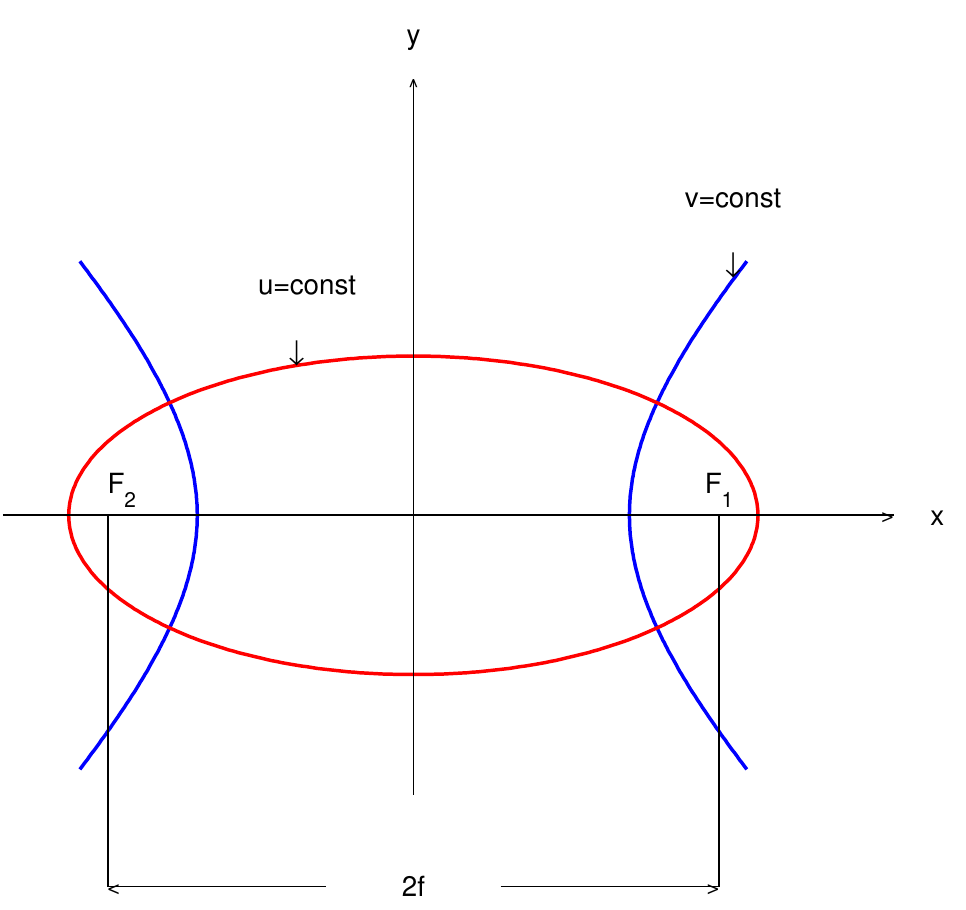}}
\caption{\label{fig:f1}The elliptic cylindrical coordinates. $F_1$ and $F_2$ are the foci of the ellipse; $f$ is the semifocal length.}
\end{center}
\end{figure}

Firstly, the elliptic cylindrical coordinates $(u,v,z)$, as shown in Fig.~\ref{fig:f1}, are introduced in terms of rectangular coordinates $(x,y,z)$
\begin{equation}
\label{eq:coord}
x = f \cosh u \,\cos v,\quad  y = f \sinh u\, \sin v,\quad  z = z ,
\end{equation}
with $0 \leq u < \infty$, $0 \leq v \leq 2\pi$, and $f$ the semifocal length of the ellipse. The contours of constant $u$ are confocal ellipses, and those of constant $v$ are confocal hyperbolas. The $z$ axis coincides with the cylinder axis. In the following we consider the transverse-electric (TE) polarized electromagnetic field (i.e., the electric field only exists in the $z$ direction). By expressing the Laplacian in elliptic cylindrical coordinates \cite{re1}, the scalar wave equation in a homogeneous dielectric medium of permittivity $\epsilon$ is given by
\begin{equation}
\label{eq:gw}
\frac{2}{\mu_0 f^2 \left(\cosh 2u- \cos 2v \right)} \Big( \frac{\partial^2 E_z}{\partial {u}^2} + \frac{\partial^2 E_z}{\partial {v}^2} \Big) + \epsilon \Big(\frac{1}{\epsilon_0 \mu_0} \frac{\partial^2 E_z}{\partial {z}^2}  - 
\frac{\partial^2 E_z}{\partial {t}^2} \Big) = 0 .
\end{equation}
An $exp(-i\omega t)$ time dependence is assumed, where $\omega$ is the circular frequency. Separation of variables implies that we assume a solution $E_z$ of the form
\begin{equation}
\label{eq:sepv}
E_z=Z(z)S(v)R(u)
\end{equation}
On substituting Eq.~\eqref{eq:sepv} in Eq.~\eqref{eq:gw} we obtain
\begin{subequations}
\begin{align}
\label{eq:sepa}
\Big( \frac{\mathrm{d}^2}{{\mathrm{d}z}^2} 
 & + k_z^2 \Big) Z(z) = 0 ,  \\
\label{eq:sepb} 
\Big[ \frac{\mathrm{d}^2}{{\mathrm{d}v}^2} 
& + \left( a - 2 q \cos 2v \right) \Big] S(v) = 0 , \\
\label{eq:sepc}
\Big[ \frac{\mathrm{d}^2}{{\mathrm{d}u}^2} 
& - \left( a - 2 q \cosh 2u \right) \Big] R(u) = 0 ,
\end{align}
\end{subequations}
where $k_z$ and $a$ are separation constants, $q=f^2k_t^2/4$, with $k_t^2=k^2-k_z^2$, where $k=k_0\sqrt{\epsilon}$, $k_0=2 \pi/\lambda$, and $\lambda$ is the wavelength of the incident light in vacuum. By taking the advantage of the elliptical cylinder geometry, we focus our analysis on the 2D scattering of normally incident plane waves $(k_z=0)$. Equation~(\ref{eq:sepb}) is known as the angular Mathieu equation. The solution is denoted by $S_{p\,m}(q,v,n)$ where $p,m$ denote even $(e)$ or odd $(o)$, and $n$ denotes the order \cite{re3,re4}. Equation~(\ref{eq:sepc}) is known as the radial Mathieu equation. Similarly to the circular cylindrical coordinates where the radial solution is expressed in terms of Bessel functions $J_n(\cdot)$, $Y_n(\cdot)$, $H_n^{(1)}(\cdot)$, and $H_n^{(2)}(\cdot)$, the radial Mathieu equation has four kinds of solutions: $J_{p\,m}$, $Y_{p\,m}$, $H_{p\,m\,1}$, and $H_{p\,m\,2}$, where $p,m = e,o$ \cite{re1,re3,re4}. 

Let a scalar plane wave of unit amplitude be normally incident on the axis of the elliptical cylinder in air, so that the propagation direction forms an angle $\phi$ with the $x$-axis in the $x-y$ plane. Hence the electric field of the incident plane wave is given by  
\begin{equation}
\label{eq:ezin0}
E_z^{in}=e^{i k_0(x\cos\phi+y\sin\phi)}=e^{i k_0 f(\cosh u
\cos v \cos \phi+\sinh u \sin v \sin \phi)}
\end{equation}
To obtain exact solutions for the scattered wave, we expand the incident field in Eq.~\eqref{eq:ezin0} in terms of the eigenfunctions of the elliptical cylinder \cite{re1,re3}
\begin{equation}
\label{eq:pw}
E_z^{in}=\sqrt{8\pi}\sum_{n}i^n J_{p\,m} (q,u,n) S_{p\,m}(q,v,n) S_{p\,m}(q,\phi ,n) / N_{p\,m}(q,n) ,
\end{equation}
where $N_{p\,m}$ is normalization constant. 

\section{Scattering by homogeneous elliptical cylinders}

Consider a homogeneous dielectric elliptical cylinder of permittivity $\epsilon_1$ with the boundary located at $u_1$, in air. The electric field $E_z$ outside and inside the elliptical cylinder is given by \cite{re2,re3}                          
\begin{align}
\label{eq:field}
(u>u_1) \quad E_z & = \sqrt{8\pi}\sum_{n}i^n J_{p\,m}(q_0,u,n) S_{p\,m}(q_0,v,n) S_{p\,m}(q_0,\phi,n)/N_{p\,m}(q_0,n)  \nonumber  \\
    & + i^n \alpha_{p\,m}^{(sc)}(n) H_{p\,m\,1}(q_0,u,n) S_{p\,m}(q_0,v,n) S_{p\,m}(q_0,\phi,n) ,  \nonumber  \\
(u<u_1)  \quad E_z & = \sqrt{8\pi}\sum_{n}i^n \alpha_{p\,m}^{(1)}(n) J_{p\,m}(q_1,u,n) S_{p\,m}(q_1,v,n) S_{p\,m}(q_0,\phi,n) ,    
\end{align}      
where $\alpha_{p\,m}^{(1)}$ and $\alpha_{p\,m}^{(sc)}$ are coefficients to be determined from the boundary conditions. The tangential magnetic field $H_v$ in each region is obtained from $E_z$ \cite{re1,re2,re3},
\begin{align}
\label{eq:hfield}
(u>u_1) \quad H_v & = \frac{1}{k_0^2 p} \Big(-i \omega \epsilon_0
\frac{\partial E_z}{\partial u} \Big) , \nonumber \\  
(u<u_1) \quad H_v & = \frac{1}{k_0^2 \epsilon_1 p} \Big(-i \omega \epsilon_1
\frac{\partial E_z}{\partial u} \Big) ,  
\end{align}      
where $p=f(\sinh^2 u + \sin^2 v)^{1/2}$. The boundary conditions require the continuity of the tangential components of the electric and magnetic field at the boundary $u=u_1$. One obtains
\begin{align}
\label{eq:bound1}
& \sum_{n}i^n [ J_{p\,m}(q_0,u_1,n)/N_{p\,m}(q_0,n) + 
\alpha_{p\,m}^{(sc)}(n) H_{p\,m\,1}(q_0,u_1,n)] S_{p\,m}(q_0,v,n) S_{p\,m}(q_0,\phi,n)   \nonumber \\
     = & \sum_{n}i^n \alpha_{p\,m}^{(1)}(n) J_{p\,m}(q_1,u_1,n) S_{p\,m}(q_1,v,n) S_{p\,m}(q_0,\phi,n) ,  \nonumber  \\
&  \sum_{n}i^n [ J_{p\,m}^\prime(q_0,u_1,n)/N_{p\,m}(q_0,n) + 
\alpha_{p\,m}^{(sc)}(n) H_{p\,m\,1}^\prime(q_0,u_1,n) ] S_{p\,m}(q_0,v,n) S_{p\,m}(q_0,\phi,n)   \nonumber \\
     = & \sum_{n}i^n \alpha_{p\,m}^{(1)}(n) J_{p\,m}^\prime(q_1,u_1,n) S_{p\,m}(q_1,v,n) S_{p\,m}(q_0,\phi,n) , \end{align}      
where the prime denotes differentiation with respect to $u$. To determine the coefficients $\alpha_{p\,m}^{(1)}(n)$ and $\alpha_{p\,m}^{(sc)}(n)$, $p,m=e,o$, we multiply Eq.~\eqref{eq:bound1} by $S_{p\,m'}(q_0,v,n)$, integrate the resulting equation over $v$ from $0$ to $2\pi$, and use the orthogonality relations of the angular Mathieu functions \cite{re1,re3,re4}. Applying those orthogonality relations allows waves in each order $n$, and each combination $p,m=e,o$ to decouple. One obtains        
\begin{align}
\label{eq:bound2}
J_{p\,m} & (q_0,u_1,n)/N_{p\,m}(q_0,n) + 
\alpha_{p\,m}^{(sc)}(n) H_{p\,m\,1}(q_0,u_1,n) \nonumber \\
& =  \alpha_{p\,m}^{(1)}(n)\gamma_{p\,m}(n) J_{p\,m}(q_1,u_1,n) /N_{p\,m}(q_0,n)
 ,  \nonumber  \\
J_{p\,m}^\prime & (q_0,u_1,n)/N_{p\,m}(q_0,n) + 
\alpha_{p\,m}^{(sc)}(n) H_{p\,m\,1}^\prime(q_0,u_1,n) \nonumber \\
& =  \alpha_{p\,m}^{(1)}(n) \gamma_{p\,m}(n) J_{p\,m}^\prime(q_1,u_1,n)  /N_{p\,m}(q_0,n), 
\end{align}      
where correlation factors $\gamma_{p\,m}(n)$, $p,m=e,o$ are defined by                         
\begin{equation}
\label{eq:gamma1}
\int_0^{2\pi}S_{p\,m'}(q_0,v,n)S_{p\,m}(q_1,v,n)\,\mathrm{d}v
=\gamma_{p\,m}(n) \delta_{m\,m'} .
\end{equation}
Expressions for the correlation factors in terms of expansion coefficients of Mathieu functions are given in \cite{re4}. From Eq.~\eqref{eq:bound2} we find 
\begin{align}
\label{eq:coeffh}
&\alpha_{p\,m}^{(1)}(n)=\frac{J_{p\,m}(q_0,u_1,n)}{\gamma_{p\,m}(n)J_{p\,m}(q_1,u_1,n)} \frac{\mathscr{J}_{p\,m}(q_0,u_1,n)-\mathscr{H}_{p\,m\,1}(q_0,u_1,n)}{\mathscr{J}_{p\,m}(q_1,u_1,n)-\mathscr{H}_{p\,m\,1}(q_0,u_1,n)} 
,  \nonumber  \\
&\alpha_{p\,m}^{(sc)}(n)=\frac{J_{p\,m}(q_0,u_1,n)}{N_{p\,m}(q_0,n)H_{p\,m\,1}(q_0,u_1,n)} \frac{\mathscr{J}_{p\,m}(q_0,u_1,n)-\mathscr{J}_{p\,m}(q_1,u_1,n)}
{\mathscr{J}_{p\,m}(q_1,u_1,n)-\mathscr{H}_{p\,m\,1}(q_0,u_1,n)} ,
\end{align} 
where we introduced the log-derivative function $\mathscr{F}=F\,^\prime \!/F$, $F$ being $J_{p\,m}$ or $H_{p\,m\,1}$, the prime denoting differentiation with respect to $u$.

\begin{figure}[h]
\begin{center}
\subfigure  {\includegraphics [height=5cm]{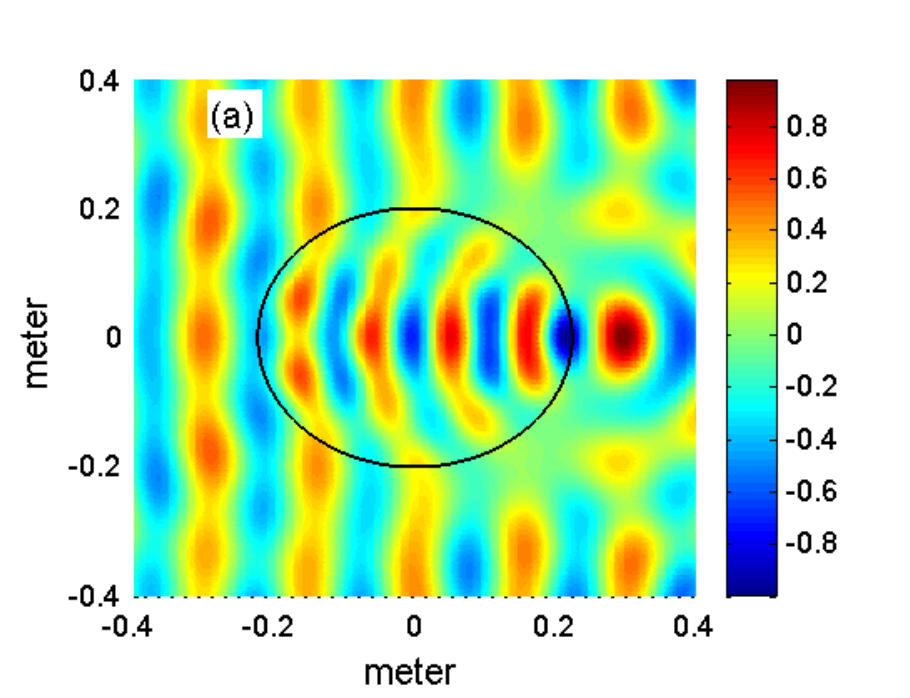}}
\subfigure  {\includegraphics [height=5cm]{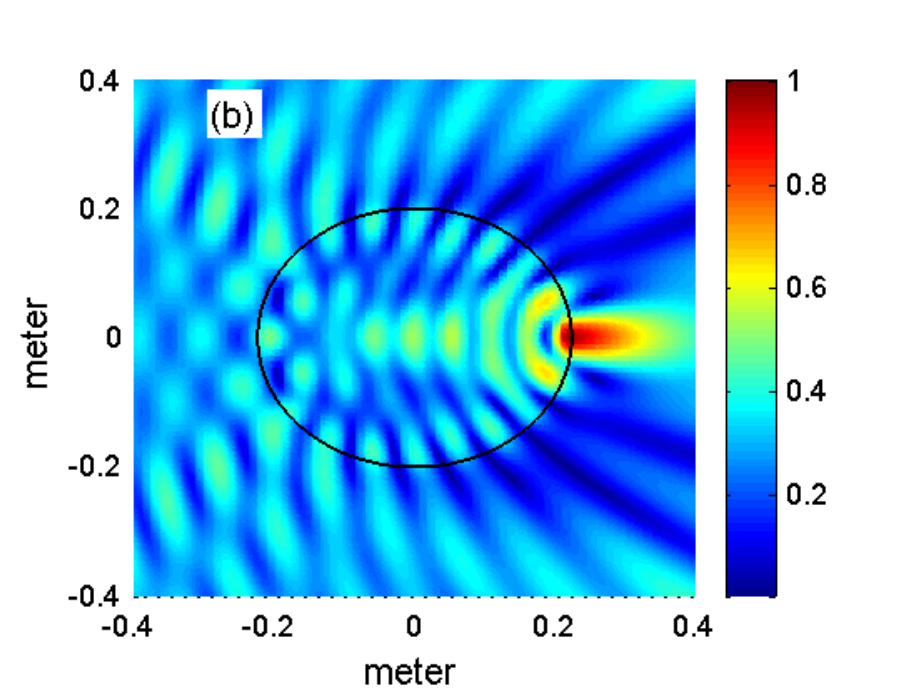}}
\subfigure  {\includegraphics [height=5cm]{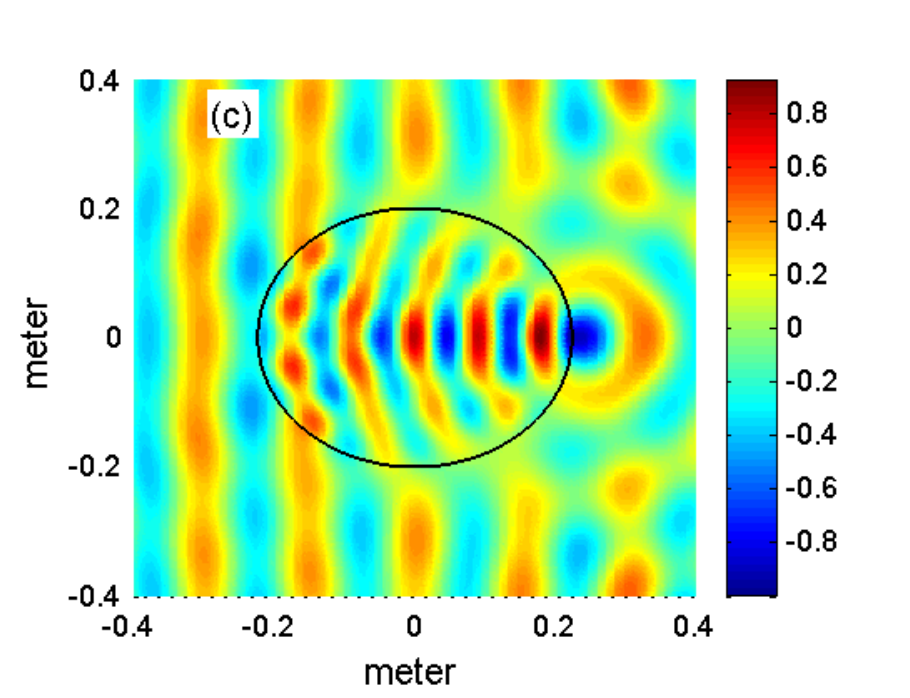}}
\subfigure  {\includegraphics [height=5cm]{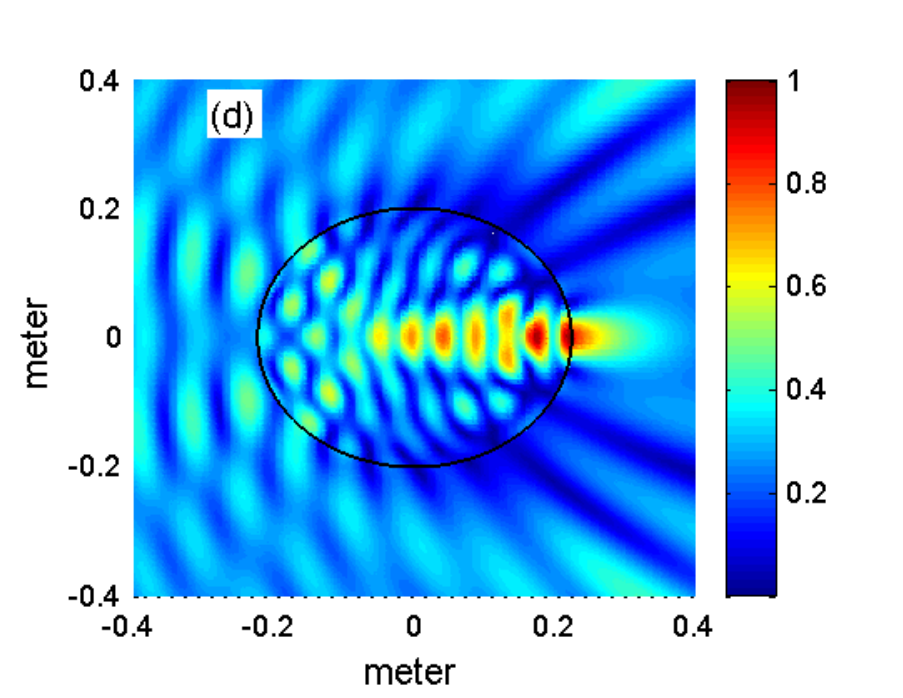}}
\caption{\label{fig:f2}(a) Snapshot and (b) norm of the resulting electric field distribution in the vicinity of an elliptical cylinder with permittivity $\epsilon=2$; (c) and (d), the same with permittivity $\epsilon=3$. The light is incident from the $-\hat{x}$ direction, in the $x-y$ plane, normally to the $z$ axis.}
\end{center}
\end{figure}

As an example we consider an elliptical cylinder of permittivity $\epsilon=2$ [Figs.~\ref{fig:f2}(a) and (b)] and $\epsilon=3$ [Figs.~\ref{fig:f2}(c) and (d)]. The semifocal length is $f=0.1$m. The elliptical boundary is at $u_1=1.4436$ (shown as black contour in the figures) with semiaxes $d_x=0.2236$m and $d_y=0.1414$m. The scatterer is illuminated from $-\hat{x}$ direction, perpendicularly to axis $z$. The incident light wavelength is $\lambda=0.15$m. We found that $n=0 \div 7$ is enough to assure the convergence of the calculated scattering field \cite{re4}. Snapshots of the normalized electric field [Real($E_z$)] distribution in the vicinity of the scatterer are shown in Figs.~\ref{fig:f2}(a) and (c) for $\epsilon=2$ and $3$, respectively, normalization being against the maximum absolute real value. Figures ~\ref{fig:f2}(b) and (d) show the distribution of $\lvert E_z \rvert$, normalized to the maximum value, for $\epsilon=2$ and $3$, respectively. One can see the focalizing effect of the perpendicularly illuminated dielectric elliptical cylinder, the greater the permittivity value, the higher the intensity on the focalizing direction. 

\section{Scattering by layered elliptical cylinders}

Consider a homogeneous dielectric elliptical cylinder of permittivity $\epsilon_1$ and boundary at $u_1$ covered by a confocal dielectric elliptical layer of permittivity $\epsilon_2$ and outer boundary at $u_2$, in air. Let $q_1$ and $q_2$ be the respective elliptical parameters, $q_1=f^2k_0^2\epsilon_1 /4$ and  
$q_2=f^2k_0^2\epsilon_2 /4$. The electric field $E_z$ in each region is given by
\begin{align}
\label{eq:field2}
(u>u_2) \quad E_z & = \sqrt{8\pi}\sum_{n}i^n [J_{p\,m}(q_0,u,n) /N_{p\,m}(q_0,n)  \nonumber  \\
    & + \alpha_{p\,m}^{(sc)}(n) H_{p\,m\,1}(q_0,u,n)] S_{p\,m}(q_0,v,n) S_{p\,m}(q_0,\phi,n) ,  \nonumber  \\
(u_1<u<u_2) \quad E_z & = \sqrt{8\pi}\sum_{n}i^n [\alpha_{p\,m}^{(2)}(n) J_{p\,m}(q_2,u,n)  \\
    & + \alpha_{p\,m}^{(3)}(n) H_{p\,m\,1}(q_2,u,n)] S_{p\,m}(q_2,v,n) S_{p\,m}(q_0,\phi,n) , \nonumber   \\
(u<u_1) \quad E_z & =\sqrt{8\pi}\sum_{n}i^n \alpha_{p\,m}^{(1)}(n) J_{p\,m}(q_1,u,n) S_{p\,m}(q_1,v,n) S_{p\,m}(q_0,\phi,n) . \nonumber   
\end{align}      
The tangential magnetic field $H_v$ in each layer is
\begin{equation}
\label{eq:hfieldj}
H_v = \frac{1}{k_0^2 \epsilon_j p} \Big(-i \omega \epsilon_j
\frac{\partial E_z}{\partial u} \Big) , \qquad j=1,2.
\end{equation}   
The boundary conditions require the continuity of the tangential components of the electric and magnetic field at the boundaries $u=u_j$, $j=1,2$. Following the same procedure as in the previous section we obtain four equations for each order $n$. (For simplicity, the dependence on order $n$ is skipped.) 
\begin{align}
\label{eq:bound3}
J_{p\,m} & (q_0,u_2)/N_{p\,m}(q_0) + 
\alpha_{p\,m}^{(sc)} H_{p\,m\,1}(q_0,u_2) \nonumber \\
& = \gamma_{p\,m}^{(2)}/N_{p\,m}(q_0)[ \alpha_{p\,m}^{(2)} J_{p\,m}(q_2,u_2) + \alpha_{p\,m}^{(3)} H_{p\,m\,1}(q_2,u_2)]
 ,  \nonumber  \\
J_{p\,m}^\prime & (q_0,u_2)/N_{p\,m}(q_0) + 
\alpha_{p\,m}^{(sc)} H_{p\,m\,1}^\prime(q_0,u_2)  \\
& = \gamma_{p\,m}^{(2)}/N_{p\,m}(q_0)[ \alpha_{p\,m}^{(2)} J_{p\,m}^\prime(q_2,u_2) + \alpha_{p\,m}^{(3)} H_{p\,m\,1}^\prime(q_2,u_2)]
 ,  \nonumber  \\ 
\gamma_{p\,m}^{(2)} & [ \alpha_{p\,m}^{(2)} J_{p\,m}(q_2,u_1) + \alpha_{p\,m}^{(3)} H_{p\,m\,1}(q_2,u_1)]=\alpha_{p\,m}^{(1)}
\gamma_{p\,m}^{(1)} J_{p\,m}(q_1,u_1) ,  \nonumber  \\ 
\gamma_{p\,m}^{(2)} & [ \alpha_{p\,m}^{(2)} J_{p\,m}^\prime(q_2,u_1) + \alpha_{p\,m}^{(3)} H_{p\,m\,1}^\prime(q_2,u_1)]=\alpha_{p\,m}^{(1)}
\gamma_{p\,m}^{(1)} J_{p\,m}^\prime(q_1,u_1) ,   \nonumber 
\end{align}      
where correlation factors $\gamma_{p\,m}^{(j)}$, $p,m=e,o$, $j=1,2$, are defined by
\begin{equation}
\label{eq:gammaj}
\int_0^{2\pi}S_{p\,m'}(q_0,v)S_{p\,m}(q_j,v)\,\mathrm{d}v
=\gamma_{p\,m}^{(j)} \delta_{m\,m'} , \quad j=1,2.
\end{equation}
Relations ~(\ref{eq:bound3}) can be written in the form $A \cdot X = B$, where
\begin{eqnarray}
\label{eq:matr}
& A = 
\begin{pmatrix}
0 \hfill & \gamma_{p \,m}^{(2)}J_{p \,m}(q_2,u_2) \hfill &\gamma_{p \,m}^{(2)}H_{p \,m\,1}(q_2,u_2) \hfill & -H_{p \,m\,1}(q_0,u_2) N_{p \,m}(q_0) \\
0 \hfill & \gamma_{p \,m}^{(2)}J_{p \,m}^\prime(q_2,u_2) \hfill &\gamma_{p \,m}^{(2)}H_{p \,m\,1}^\prime(q_2,u_2) \hfill & -H_{p \,m\,1}^\prime(q_0,u_2) N_{p \,m}(q_0) \\
-\gamma_{p \,m}^{(1)}J_{p \,m}(q_1,u_1) \hfill & \gamma_{p \,m}^{(2)}J_{p \,m}(q_2,u_1) \hfill & \gamma_{p \,m}^{(2)}H_{p \,m\,1}(q_2,u_1) \hfill & 0 \\
-\gamma_{p \,m}^{(1)}J_{p \,m}^\prime(q_1,u_1) \hfill & \gamma_{p \,m}^{(2)}J_{p \,m}^\prime(q_2,u_1) \hfill & \gamma_{p \,m}^{(2)}H_{p \,m\,1}^\prime(q_2,u_1) \hfill & 0
\end{pmatrix} ,  \nonumber \\
& X = 
\begin{pmatrix}
\alpha_{p\,m}^{(1)} \\  \alpha_{p\,m}^{(2)} \\ 
\alpha_{p\,m}^{(3)} \\ \alpha_{p\,m}^{(sc)} 
\end{pmatrix},  \quad
\textrm{and} \quad
B=
\begin{pmatrix}
J_{p\,m}(q_0,u_2) \\ J_{p\,m}^\prime(q_0,u_2) \\ 
 0 \\ 0 
\end{pmatrix} .
\end{eqnarray}
The unknown coefficients $\alpha_{p\,m}^{(j)}$ ($j=1,2,3,sc$) are determined at each order $n$ by solving the matricial equation. In 
\textsc{matlab}, an equation of this form is solved simply with command $X=A \backslash B$.

\begin{figure}
\begin{center}
\subfigure  {\includegraphics [height=5cm]{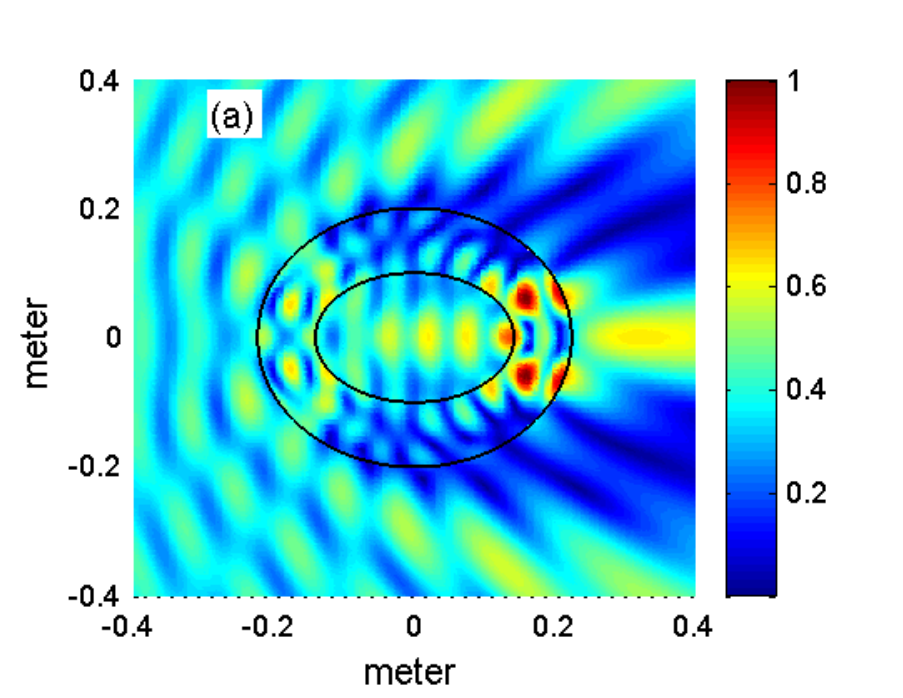}}
\subfigure  {\includegraphics [height=5cm]{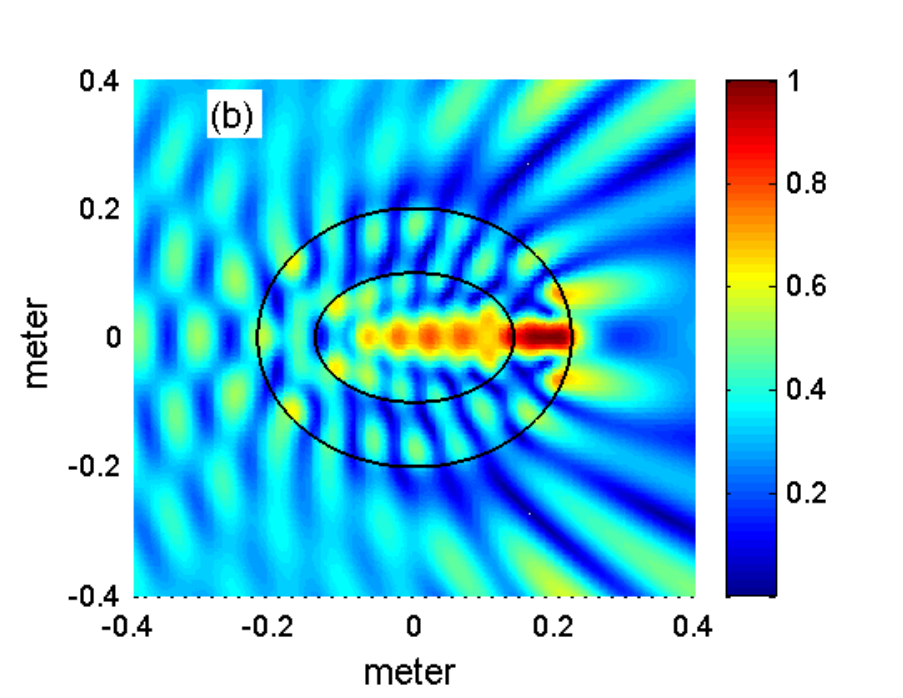}}
\caption{\label{fig:f3}(a) Norm of the resulting electric field distribution in the vicinity of a layered elliptical cylinder with inner (outer) permittivity $\epsilon_1=2 (\epsilon_2=3)$ and (b)  with inner (outer) permittivity $\epsilon_1=3 (\epsilon_2=2)$. The light is incident from the $-\hat{x}$ direction, in the $x-y$ plane, normally to the $z$ axis.} 
\end{center}
\end{figure}

As an example, we show the distribution of $\lvert E_z \rvert$ normalized to the maximum value for a layered elliptical cylinder with permittivities $\epsilon_1=2$, $\epsilon_2=3$ in Fig.~\ref{fig:f3}(a), and $\epsilon_1=3$, $\epsilon_2=2$ in Fig.~\ref{fig:f3}(b). The semifocal length is $f=0.1$m. The inner boundary is at $u_1=0.8814$ (with semiaxes $d_x=0.2$m, $d_y=0.1$m) and the outer boundary is at $u_2=1.4436$ (with semiaxes $d_x=0.2236$m, $d_y=0.1414$m), the boundaries being shown as black contours in the figure. The wavelength of the incident light is $\lambda=0.15$m, the light being incident from the $-\hat{x}$ direction, normally to the $z$ axis. One can see that a confocal cover influences on the scattered field distribution; a lower (higher) permittivity cover decreases (increases) the intensity of the scattered field. The focalizing effect of the perpendicularly illuminated elliptical cylinder is increased by a lower permittivity confocal cover. 

\section{Summary}

In this contribution we have studied the plane wave scattering by homogeneous and layered elliptical dielectric cylinders following an analytical approach that is based on Mathieu functions. By taking the advantage of the elliptical cylinder geometry, we focused our analysis on the 2D scattering of normally incident plane waves. We have provided closed-form relations for the expansion coefficients of the resulting electric field in the vicinity of the homogeneous and layered dielectric elliptical cylinders. Numerical examples have been provided showing the focalizing effect of a homogeneous elliptical cylinder illuminated perpendicularly to the axis, and the influence of a confocal cover layer on the resulting electric field distribution. The results presented could be a valuable contribution to the enlargement of the Mathieu functions applicability.

\newpage

\begin{thebibliography}{4}

\bibitem{re1}J. A. Stratton, \textsl{Electromagnetic Theory} (Mc-Graw Hill, New York, 1941).

\bibitem{re2}C. Yeh, J. Math. Phys. {\bf 4}, 65 (1963).

\bibitem{re3}J. J. Stamnes and B. Spjelkavik, Pure Appl. Opt. {\bf 4}, 251 (1995); J. J. Stamnes {\it ibid.} {\bf 4}, 841 (1995).

\bibitem{re4}E. Cojocaru, \textsc{matlab} free available computer code \href{http://www.mathworks.com/matlabcentral/fileexchange/}
{Mathieu Functions Toolbox v. 1.0}.

\end{thebibliography}
\end{document}